\documentclass[multphys,vecphys]{svmult}

\usepackage{makeidx}     
\usepackage{graphicx}    
\usepackage{multicol}    

\makeindex             

\begin{document}

\title*{Second Epoch Global VLBI Observations of Compact Radio Sources in the M82 Starburst Galaxy}
\titlerunning{Global VLBI Observations of Compact Radio Sources in M82}
\author{J.~D.~Riley\inst{1}\and
A.~Pedlar\inst{1}\and T.~W.~B.~Muxlow\inst{1}\and A.~R.~McDonald\inst{1}\and R.~J.~Beswick\inst{1}\and K.~A.~Wills\inst{2}}
\authorrunning{J.~D.~Riley et al.}
\institute{Jodrell Bank Observatory, University of Manchester, Macclesfield, Cheshire. SK11 9DL. UK
\texttt{jriley@jb.man.ac.uk}
\and Department of Physics and Astronomy, University of Sheffield, Sheffield. S3 7RH. UK}
%
%
\maketitle

\abstract{We have presented the results of a second epoch of global Very Long Baseline Interferometry observations, taken on 23 February 2001 at a wavelength of $18 \, \mathrm{cm}$, of the central kiloparsec of the nearby starburst galaxy Messier 82. These observations were aimed at studying the structural and flux evolution of some of the compact radio sources in the central region that have been identified as supernova remnants. The objects 41.95+575 and 43.31+592 have been studied, expansion velocities of $2500 \pm 1200 \, \mathrm{km} \, \mathrm{s}^{-1}$ and $7350 \pm 2100 \, \mathrm{km} \, \mathrm{s}^{-1}$ respectively have been derived. Flux densities of $31.1 \pm 0.3 \, \mathrm{mJy}$ and $17.4 \pm 0.3 \, \mathrm{mJy}$ have been measured for the two objects. These results are consistent with measurements and predictions from previous epochs.}

\section{Introduction}
\label{intro}
The nearby ($3.2 \, \mathrm{Mpc}$ distant) starburst galaxy Messier 82 (M82) has a high inferred star formation rate of~$\simeq 1.7~\mathrm{to}~2.2 \, \mathrm{M}_{\odot} \, \mathrm{yr}^{-1}$ for stars with masses greater than $5 \, \mathrm{M}_{\odot}$ (\cite{pedlar01})which leads to an expected supernova rate of $0.05~\mathrm{to}~0.1\, \mathrm{yr}^{-1}$ . As a result M82 is an ideal laboratory for the study of radio supernovae in a starburst environment and such studies can yield important information on the properties of the interstellar medium (ISM).
The central region of M82 hosts in excess of 60 compact radio sources, most of which have now been identified as shown in \cite{mcdonald02} and the references therein. Of these sources 16 have been identified as \textsc{Hii} regions with at least 30 of the others identified as supernova remnants (SNR). Several studies have been conducted of these compact sources. \cite{mcdonald02} used the Multi Element Radio Linked Interferometer Network (MERLIN) to study the spectral indices of the sources to aid in identification. Others have concentrated on the SNR studying them at various angular scales such as \cite{muxlowthis} using MERLIN and two using Very Long Baseline Interferometry (VLBI) techniques: \cite{pedlar99} using the European VLBI Network (EVN) and \cite{mcdonald01} along with \cite{pedlarthis} and this paper using a global VLBI network. These studies have primarily been focused at measuring the expansion velocities and deceleration parameters of the SNR and thereby testing the models put forward in \cite{chevalier01}.

\section{Observations and Data Analysis}
\label{obs}
The observations were made on 23 February 2001 using a 19 station global VLBI network at a wavelength of $18 \, \mathrm{cm}$. The network consisted of eleven antennas of the Very Long Baseline Array (VLBA) in the USA, 6 antennas from the EVN, the NASA Deep Space Network (DSN) antenna at Robledo, Spain and a single Very Large Array (VLA) dish. This network was virtually identical to the 20 station network used by \cite{mcdonald01} on 28 November 1998, with only the NASA DSN antenna at Goldstone, USA missing compared to the first epoch. The data were taken in spectral line mode using 128 channels each of bandwidth $0.125 \, \mathrm{MHz}$, yielding a total bandwidth of $16 \, \mathrm{MHz}$.

Observations were made of several sources, the target source being the central kiloparsec of M82, with J0958+65 being observed throughout the run as a phase calibration source. The sources 3C84 and J0927+39 were used for fringe fitting and bandpass calibration, with 3C84 alone being used for flux scale calibration.

The data were acquired and correlated using the Mark IV/VLBA system and the data reduction and image analysis were conducted using the Astronomical Image Processing System (\textsc{aips}) which is provided and maintained by the National Radio Astronomy Observatory (NRAO) in the USA.
\section{Compact Source: 41.95+575}
\label{4195sec}
\begin{figure}[!h]
\centering
\includegraphics[height=5cm]{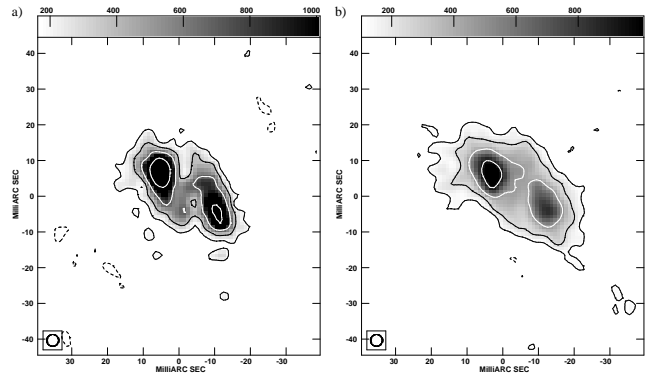}
%
%
\caption{Maps of 41.95+575 from both epochs of the global VLBI observations convolved with a $3.5 \, \mathrm{mas}$ circular beam. a)~Epoch 28 November 1998 map with contours at $(1, 2, 4, 8) \times 173.6 \, \mu \mathrm{Jy} \, \mathrm{beam}^{-1}$. b)~Epoch 23 February 2001 map with contours at $(1, 2, 4, 8) \times 114.4 \, \mu \mathrm{Jy} \, \mathrm{beam}^{-1}$.}
\label{4195map}       
\end{figure}
The source 41.95+575 (Figure \ref{4195map}) is the most compact of the sources in M82 and does not exhibit a shell structure, but appears to be more bipolar in nature. Difficulties were encountered when trying to measure the expansion of this object. The angular expansion velocity is quite low and leads to an expected expansion of about $1 \, \mathrm{mas}$ in the time between the two epochs. The other problem is that changes in the small scale structure make finding common reference points between epochs to measure the expansion difficult.

The most reliable method of making expansion measurements in this source was found to be moment fitting. This was conducted using the \textsc{aips} task \textsc{momft}. Using this method on data from both epochs we arrive at sizes and other parameters for the source mentioned in Table \ref{4195tab}. From these results we calculate a total angular expansion of $0.77 \pm 0.37\, \mathrm{mas} \, \mathrm{yr}^{-1}$ along the major axis. The expansion along the major axis has been  measured in the past \cite{trotman96}, \cite{mcdonald01} and more recently in \cite{muxlowthis}. The results presented here are both consistent with those from \cite{trotman96} and \cite{muxlowthis}. Using the standard distance to M82 of $3.2 \, \mathrm{Mpc}$ it is possible to calculate that the linear expansion speed from the centre, along the major axis, perpendicular to the line-of-sight is $2500 \pm 1200 \, \mathrm{km} \, \mathrm{s}^{-1}$. It is necessary to note that as this source appears to be bipolar in nature the expansion along the major axis may be greater than the value quoted here due to orientation effects.
\begin{table}
\centering
\caption{The sizes of the source 41.95+575 calculated using moment fitting for both of the recent global VLBI epochs.}
\label{4195tab}       
%
%
\begin{tabular}{cccc}
\hline\noalign{\smallskip}
Epoch & Major Axis & Error & Position Angle \\
 & (mas) & (mas)  & (degrees) \\
\noalign{\smallskip}\hline\noalign{\smallskip}
28 Nov 1998 &  $21.4$ & $\pm 0.8$ & 54.63 \\
23 Feb 2001 & $23.1$ & $\pm 1.4$ & 58.82 \\
\noalign{\smallskip}\hline
\end{tabular}
\end{table}

In the 2.2 years between the observations the flux density of 41.95+575 has decreased from $43.8 \pm 2.3 \, \mathrm{mJy}$ to $31.1 \pm 0.3 \, \mathrm{mJy}$. This in line with expectations that the source continues to decrease at a rate of $8.5 \, \%$ per year.
\section{Compact Source: 43.31+592}
\label{4331sec}
\begin{figure}[!h]
\centering
\includegraphics[height=5cm]{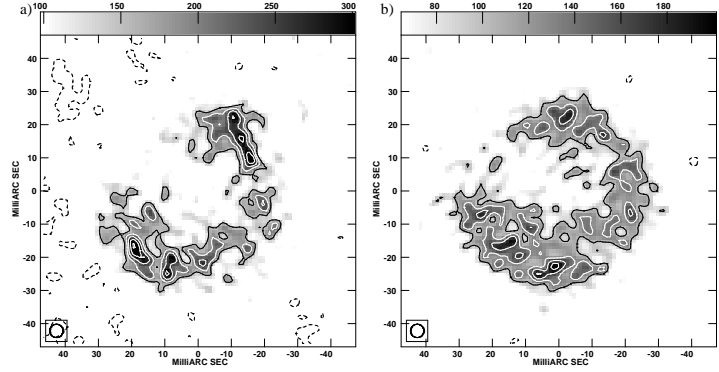}
%
%
\caption{Maps of 43.31+592 from both epochs of the global VLBI observations convolved with a $4 \, \mathrm{mas}$ circular beam. a)~Epoch 28 November 1998 map with contours at $(1, 1.5, 2, 2.5 3) \times 100 \, \mu \mathrm{Jy} \, \mathrm{beam}^{-1}$. b)~Epoch 23 February 2001 map with contours at $(1, 1.5, 2, 2.5 3) \times 66 \, \mu \mathrm{Jy} \, \mathrm{beam}^{-1}$.}
\label{4331map}       
\end{figure}
\begin{figure}[!t]
\centering
\includegraphics[height=5cm]{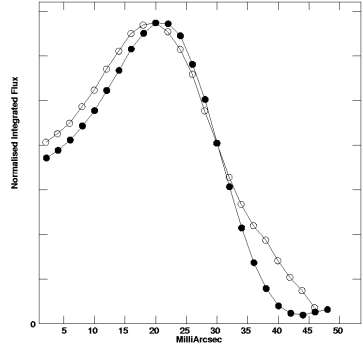}
%
%
\caption{Results of integrating around annuli from the centre of the SNR 43.31+592. The results for 2001 epoch shown by filled circles and for 1998 epoch by ufilled circles}
\label{4331iring}       
\end{figure}

The compact source 43.31+592 (Figure~\ref{4331map}) is a typical shell-type SNR. The expansion measurements of this source posed problems due to evolution of the small-scale structure which did not allow accurate Gaussian fitting to the four knots of emission identified in \cite{mcdonald01}. Instead, the centre of the source in the maps was found for each epoch and the flux in $2 \, \mathrm{mas}$-thick annuli was integrated around each annulus using the \textsc{aips} task \textsc{iring}. A plot of the integrated flux in each annulus against the distance from the centre was made (see Figure \ref{4331iring}). The position of the peaks for each epoch were then used to calculate the expansion speed of the SNR. The 1998 data exhibit increasing errors towards the outer edge of the SNR due to confusion between diffuse emission and the background noise.
The peak for the 1998 epoch appears at a radius of $19 \, \mathrm{mas}$ and in the 2001 epoch the peak appears at $21.5 \, \mathrm{mas}$. As a result the angular expansion speed is $1.13 \pm 0.32\, \mathrm{mas} \, \mathrm{yr}^{-1}$ which corresponds to a linear expansion speed of $7350 \pm 2100 \, \mathrm{km} \, \mathrm{s}^{-1}$. These preliminary results are  consistent with the results in both \cite{mcdonald01}, \cite{pedlar99} and \cite{muxlowthis}. The flux density of the source has changed from $19.5 \pm 0.1 \, \mathrm{mJy}$ during the 1998 epoch to $17.4 \pm 0.3 \, \mathrm{mJy}$ during the 2001 epoch.
\section{Conclusion}
Two of the compact radio sources in M82 that have been identified as SNR have been studied at two epochs with global VLBI techniques.

The source 41.95+575 is expanding at a speed of  $2500 \pm 1200 \, \mathrm{km} \, \mathrm{s}^{-1}$, which is consistent with previous measurements. The flux continues to decline at a rate of 8.5 percent per year and during the latest epoch stood at $31.1 \pm 0.3 \, \mathrm{mJy}$.

43.31+592 is expanding at a speed of $7350 \pm 2100 \, \mathrm{km} \, \mathrm{s}^{-1}$, which continues to support the proposal \cite{mcdonald01} that the SNR is in or close to free expansion. This result is in conflict with models proposed in \cite{chevalier01}.

The ultimate goal of better constraining the expansion velocities and measuring the deceleration parameters for these SNR can only be achieved by further observations at this resolution. In order to allow for enough expansion to occur to allow detection it has been suggested that this will be done every $4~\mathrm{to}~5$ years.
%

%


\printindex
\end{document}